 \documentclass[preprint2]{aastex}

\usepackage{graphicx}
\usepackage{longtable}
\usepackage{graphicx}
\usepackage{longtable}
\usepackage{natbib}
\usepackage{color}
\usepackage{amssymb, amsmath}
\def\smbh{SMBH\ }
\def\hess{H.E.S.S.\ }

\def\mbh{M_{\mathrm{BH}}}
\def\Mrid{M_{9}}

\def\gr{$\gamma$-ray\ }
\def\grs{$\gamma$-rays\ }

\def\ngjm{n_{\mathrm{GJ}}}

\newcommand\simgt{\lower.5ex\hbox{$\; \buildrel > \over \sim \;$}}
\newcommand\simlt{\lower.5ex\hbox{$\; \buildrel < \over \sim \;$}}

\slugcomment{Not to appear in Nonlearned J., 45.}

\shorttitle{VHE \gr emission from Passive
  SMBH: NGC 1399}
\shortauthors{Pedaletti et al.}

\begin{document}


\title{VHE \gr Emission from Passive Supermassive Black Holes: Constraints for NGC 1399}

\author{G. Pedaletti\altaffilmark{1} and S. J. Wagner}
\affil{Landessternwarte, Universit\"at Heidelberg, K\"onigstuhl, D 69117 Heidelberg, Germany}
\email{gpedalet@lsw.uni-heidelberg.de}

\author{F.M. Rieger\altaffilmark{2}}
\affil{Max-Planck-Institut f\"ur Kernphysik, P.O. Box 103980, D 69029
Heidelberg, Germany}

\altaffiltext{1}{now at Institut de Ciencies de L'Espai (IEEC-CSIC), Campus UAB, Fac. de Ci\'{e}ncies, Torre C5, parell, 2a planta 
}
\altaffiltext{2}{European Associated Laboratory for Gamma-Ray Astronomy, jointly supported by CNRS and MPG}

\begin{abstract} 
  Very high energy (VHE; $>$100 GeV) $\gamma$-rays are expected to be emitted from the 
  vicinity of super-massive black holes (SMBH), irrespective of their activity state. In the 
  magnetosphere of rotating SMBH, efficient acceleration of charged particles can take place 
  through various processes. These particles could reach energies up to $E\sim10^{19}$eV. 
  VHE \gr emission from these particles is then feasible via leptonic or hadronic processes. 
  Therefore passive systems, where the lack of a strong photon field allows the VHE \grs
  to escape, are expected to be detected by Cherenkov telescopes. We present results
  from recent VHE experiments on the passive SMBH in the nearby elliptical galaxy NGC 
  1399. No \gr signal has been found, neither by the \hess experiment nor in the Fermi data 
  analyzed here. We discuss possible implications for the physical characteristics of the 
  system. We conclude that in a scenario where particles are accelerated in vacuum gaps 
  in the magnetosphere, only a fraction $\sim 0.3$ of the gap is available for particle acceleration, 
  indicating that the system is unlikely to be able to accelerate protons up to $E\sim10^{19}$eV. 
\end{abstract}


\section{Introduction}
Spheroidal systems (such as elliptical galaxies, lenticular galaxies, and early-type spiral galaxies 
with massive bulges) are commonly believed to host super-massive black holes with masses in 
the range $\mbh= (10^6 - 10^9)~M_\odot$ in their central region \citep[e.g.,][]{rich,ferr-ford}. Many 
SMBH are {\it active} and give rise to emission over a wide range of frequencies. Some have been 
detected from radio to high energy \grs ($\sim1$ GeV). The SMBH is essential for the production 
and stability of relativistic jets. In many objects the inner, compact jets emit VHE \gr emission 
through various leptonic and/or hadronic processes. In blazar-type objects, where the jet is pointing 
towards the observer, VHE \gr emission is frequently observed \citep[e.g.,][]{hin-hof,weekes}. Often, 
this can be successfully modeled by non-thermal electrons upscattering either their own synchrotron 
radiation or upscattering an external photon field, although a hadronic origin \citep[e.g.,][]{aharonian00,
muc-prot} cannot be excluded. In these cases the gamma-ray emission is released through the jets 
which in turn are powered by the SMBH.

VHE emission may also be indirectly related to SMBH  independent of any electromagnetic activity. 
The presence of a SMBH steepens the potential well and hence the density profile of dark matter in 
the central regions of galaxies. The rate of annihilation of Dark Matter particles \citep[e.g.][]{darkmatter} 
will thus be increased resulting in enhanced emission of the gamma-rays generated in the annihilation.

Several models for the direct production of \gr emission in the vicinity of SMBH have been proposed 
\citep[see e.g.,][]{cascadeinagn,mastichiadis,slane,boldt-gosh,lev,neron,neron-aha,rieger,istomin-sol,osm10,lev11}. 
Possibly, some of these mechanisms could also be responsible for acceleration of cosmic-rays to energies 
$E\sim10^{19}$ eV and beyond. In order to avoid attenuation on circumnuclear fields, VHE photons could 
escape only if the SMBH does not produce too much low-energy radiation. The SMBH would have to be 
passive at low energies, i.e. most of the radiative losses would have to occur at high energies. In all cases 
a large mass of the central object is an important characteristic for generating a high VHE flux.

Correlations involving the SMBH masses and properties of their host galaxies have been investigated 
by many authors. In particular $\mbh$ is found to be linked to the central stellar velocity dispersion 
\citep[e.g.,][]{gebhardt} or to the mass of the host galaxy bulge \citep[e.g.,][]{magorrian}.  These
observational scaling laws, in addition to confirming the ubiquity of SMBH, have suggested that their 
activity  is tightly linked to the evolution of their host galaxies \citep[e.g.][]{ferr-ford}.  During the 
early stages of galaxy evolution SMBH accrete matter at high rates and are observed as bright QSO 
(Quasi-Stellar Objects). Even if such systems generate VHE gamma-rays, the dense photon fields 
associated with the Quasar phase would pair-absorb the VHE radiation. The average radiative output 
at low photon energies (e.g., in the optical band) decays from redshift $z>3$ to $z=0$ by almost 2 orders of 
magnitude. The majority of SMBH in the local universe are hosted in systems of low accretion rate and 
are therefore not embedded in dense radiation fields. In {\it passive} systems (i.e., SMBH hosted in nuclei 
without bright signatures of broad-band activity and very low luminosity in longer wavelengths). If VHE 
gamma-rays, if generated, can escape from the nuclear region without suffering from strong absorption 
via photon-photon pair absorption.

While the detection of VHE gamma-rays in blazar-type systems is facilitated by the superluminal motion 
(the apparent luminosity is boosted and the optical depth related to photon-photon absorption is reduced, 
see \citet{blaz_sch}), this is not the case for non-blazar systems. Hence, proximity and low luminosity
in the IR/optical domain increase the possibility of a detection in the VHE band. Observations of passive 
systems can hence contribute to our understanding of the physics and properties of galactic nuclei. In 
addition they might give us an insight on Ultra High Energy Cosmic Ray (UHECR; $E>4 \times 10^{18}$ 
eV) sources.

Here we present GeV limits and discuss observations of the passive SMBH in the core of NGC 1399, the 
central galaxy of the Fornax cluster, that were conducted with the \hess Cherenkov telescope array 
\citep{mio_icrc} and discuss its implications for gap-type particle acceleration and emission models.

\section{The test object NGC 1399}
The most massive black holes can be found in the large ellipticals at the center of galaxy groups and clusters.
The nearest \smbh reside in the giant radio galaxy Centaurus~A and in the central region of the Virgo cluster, 
especially in M87. These galactic nuclei have already been detected in VHE observations by, e.g., the \hess 
experiment. The detected VHE emission might well be associated with features of their powerful inner jets 
\citep{m87,cena}. Also SgrA*, in the galactic center, would be an obvious good candidate, despite its low black 
hole mass of $\mbh\sim4\times10^6 M_\odot$, thanks to its proximity. However, it is very close to other
established Galactic gamma-ray emitting sources. Hence, these galaxies do no represent the best candidates 
for the studies described above. 

The most nearby cluster visible from the southern hemisphere is the Fornax cluster. It was observed by \hess 
\citep{mio_icrc}. The giant elliptical galaxy NGC 1399 is located in the central region of the Fornax cluster at 
a distance of $20.3$ Mpc. A SMBH of $ \mbh = 5.1\times10^8M_\odot$ resides in its central region 
\citep{gebhardt_mas}. The nucleus of this galaxy is well known for its low emissivity at all wavelengths, e.g., 
see \cite{oconn} and references therein.
 
The galaxy shows low-power antiparallel jets, with a luminosity of $\sim10^{39} \mathrm{erg \ s^{-1}}$ between 
$10^7$ Hz and $10^{10}$ Hz \citep{killeen}. The outflows are confined in projection within the optical extension 
of the galaxy. The jets are initially transonic, then decelerate in the inter-stellar medium and end in lobe-like 
diffuse structure at $\sim9 $ kpc from the center \citep{killeen}. X-ray images reveal cavities suggesting that 
the radio-emitting plasma is producing shocks in the hot X-ray emitting gas \citep{shurkin}. The estimated jet
power is $L_{\mathrm{jet}} \simeq 10^{42} \textrm{ erg s}^{-1}$. 

Given the suggested low photon density at low energies, the high mass of the central SMBH, the absence of 
luminous jets and  proximity of the galaxy, NGC 1399 emerges as the best candidate for a study of passive 
SMBH at very high energies.

\section{Estimate of VHE luminosity from Passive SMBH}\label{ch:theo}
The expected luminosity and energy range of VHE \gr emission from magnetospheric processes in SMBH
can be estimated as follows:

If the central black hole is spinning and accretes matter that carries magnetic flux, it will develop a rotating 
magnetosphere. This magnetosphere may be similar to those of neutron stars. Ambient charged particles 
can be accelerated to very high energies in the magnetosphere and radiative cooling will result in high 
energy radiation. A full treatment of the acceleration and of the resultant VHE \gr spectra requires detailed 
modelling of charged particle trajectories and interactions, as shown in, e.g., \cite{neron-aha}. 

It is however possible to estimate the expected VHE \gr luminosity based on some simple arguments: \\
The maximum available potential due to field line rotation is $\Delta V \simeq (a/2)~r_H B$ \citep{thorne}, 
where $a=J/J_{\mathrm{max}}$ (with $J_{\mathrm{max}}=GM^2/c$ the maximum angular momentum) is 
the black hole spin parameter, $r_H$ denotes the radius of the event horizon and $B$ the strength of the 
ordered magnetic field component. Provided a gap of size $h$ exists, the effective potential for particle 
acceleration is (Levinson 2000)
\begin{equation}\label{V_effective}
  \Delta V_e \simeq \Delta V (h/r_H)^2\,.
\end{equation} 
If particles are accelerated along the magnetic field lines, the primary radiative loss is curvature emission (note that photo-meson production is highly
inefficient in low ambient radiation fields):
\begin{equation}\label{eq:curvature_loss}
 P_{\mathrm{curv}}= \frac{2}{3}\frac{q^2c}{R_\mathrm{curv}^2}\gamma^4,
\end{equation}
with $q$ the particle charge, $R_\mathrm{curv}$ the curvature radius and $\gamma$ the Lorentz factor of 
the particle. Balancing acceleration in the gap by losses implies
\begin{equation}
q~\Delta V_e = P_{curv} h/c\,.
\end{equation} 

The expected emission spectrum from curvature radiation then extends up to VHE energies and the energy 
of the curvature photons does not depend on the mass of the emitting particle, i.e.,  
\begin{eqnarray}\label{eq:cutoff}
 E_{\mathrm{\gamma,curv}} &=& \frac{3 \hbar c \gamma^3}{2 R_{\rm curv}} 
                         \simeq 4.8~a^{3/4}\Mrid^{1/2} B_4^{3/4}\nonumber\\
                         &&\times \left(\frac{h}{r_H}\right)^{3/4}\left(\frac{R_\mathrm{curv}}{r_g}\right)^{1/2}
                         \mathrm{TeV,}
\end{eqnarray} where $M_9=\mbh/10^9 M_{\odot}$, $B_4=B/10^4$ G, and $r_g=1.5\times 10^{14} M_9$ 
cm is the gravitational radius. Hence, for realistic magnetic field strengths ($B>>1$G), curvature emission
is expected to peak in the Fermi and/or H.E.S.S. energy range. In the case of curvature-limited losses, an 
accelerated proton could reach (provided the potential is large enough) ultra-high energies of 
\begin{equation}\label{eq:cutoff_part}
 E_{\mathrm{p,curv}}\simeq2 \times 10^{19}a^{1/4}\Mrid^{1/2}B_4^{1/4}
                                      \left(\frac{hR_\mathrm{curv}^2}{r_H^3}\right)^{1/4} \textrm{eV.}
\end{equation}

In terms of the Goldreich-Julian number density $n_{\rm GJ}=\Omega B \cos \theta/(2\pi c q)$ \citep{gold-jul}, 
the maximum VHE power of the gap for $(n_e/n_{\rm GJ}) \leq 1$ is given by
\begin{eqnarray}
 L_{\gamma} &=& \int n_e~P_{\rm curv}~dV\\ \nonumber 
                        &\simeq& \eta \left(\frac{n_e}{\ngjm}\right)\frac{\Omega_H B}{2\pi q c}~(q \Delta V_e)~c r_H^2\,,
\end{eqnarray} where $\Omega=a~(c/2r_H)$ is the angular frequency of the black hole and $\eta \simlt 1$ 
is a geometrical factor, and where the volume element has been approximated by $dV\simeq \pi h r_H^2$.
Using Eq.~(\ref{V_effective}) one finally obtains
\begin{eqnarray}\label{eq:power}
 L_{\gamma} &\simeq& \frac{\eta}{8} a^2 \left(\frac{n_e}{\ngjm}\right) B^2 \left(\frac{h}{r_g}\right)^2 r_g^2 c \nonumber \\
&\simeq& 8 \times 10^{45} \eta~a^2 \left(\frac{n_e}{\ngjm}\right) B_4^2~M_9^2 \left(\frac{h}{r_g}\right)^2\nonumber\\ 
 &      & \textrm{erg s}^{-1}\,.
\end{eqnarray} 
Note that for $n_e \sim n_{\rm GJ}$, $L_{\gamma}$ is roughly a fraction $(h/r_g)^2$ of the maximum 
Blandford-Znajek jet power $L_{\rm BZ}$.

In the case of NGC 1399, the black hole mass is $M_9\simeq0.5$. The spin parameter $a$ depends on the 
formation history of the black hole. If its growth is dominated by gas accretion and the accreted mass is of 
the order of the initial mass of the black hole, then it is possible to reach spins approaching $a\simeq1$
\citep[see e.g.,][]{volonteri_spin}. This is believed to be true for the SMBH hosted in elliptical galaxies in 
the nearby universe, but might be violated if chaotic accretion occurs \citep{king08}.\\
A lower limit on the magnetic field ($B \sim 10$ G) in the inner disk might be obtained by assuming 
equipartition with the observed radiation fields. However, as the radiative output is very low, the disk is
unlikely to be of the standard-type. Thus, a more realistic value is given by equipartition with the accretion 
energy density, i.e., $B^2/8\pi = 1/2\rho(r_0)v^2_r(r_0)$, where $\rho$ is the mass density and $v_r$ is 
the radial infall velocity. The plasma density at the Schwarzschild radius is $\rho(r_s) \simeq 3 \times 
10^{-16}$ g/cm$^3$, when extrapolated from the value at the accretion radius \citep{pellegrini}, assuming 
a free-fall profile. With $v_r(r_s)=c/\sqrt{2}$, this would suggest a field strength $B(r_s)\sim 1300$ G. A 
similar value is obtained if the inner disk would be of the ADAF/radiatively inefficient accretion flow-type 
\citep[]{narayan98}. In fact, due to its low overall luminosity NGC~1399 has become a prominent ADAF 
candidate source \citep[e.g.,][]{narayan02}. One thus expects the accretion rate in NGC~1399 to be close 
to $\dot{m}\simeq 10^{-4} \dot{m}_{\rm Edd}$ \citep[cf.][]{lowen,narayan02} and the inner disk magnetic 
field to be close to $B_4 \simeq 0.1$. Note that these values are compatible with the requirement that the 
jet power ($L_{\rm jet} \simeq 10^{42}$ erg/s) is ultimately provided by a Blandford-Znajek-type process.

The above suggests that the VHE luminosity in NGC 1399 could be as high as
\begin{equation}\label{eq:luminosity_adaf}
     L_\gamma \simeq 10^{43} a^2  \left(\frac{n_e}{\ngjm}\right) \left(\frac{h}{r_g}\right)^2 \quad 
     \mathrm{ erg~s}^{-1}\,,
\end{equation} and hence high enough to be detectable by a VHE telescope array like the \hess experiment.

\section{Radiation Fields of NGC 1399}
In constructing the spectral energy distribution (SED) of NGC~1399, observations were selected in order to 
consider only its nuclear region. To achieve this, at any wavelength only the observations taken with the 
smallest aperture reported in the literature were included (see Fig. \ref{fig:SED}). In most energy bands the 
smallest apertures exceed the projected linear scales considered above by several orders of magnitude. 
The measurements are hence to be considered as upper limits of the flux emitted from the vicinity of the SMBH. 

\subsection{Gamma-ray observations}

In the VHE band NGC 1399 has been observed with the H.E.S.S. instrument \citep{mio_icrc,mio_gamma08}. 
The observations did not result in a detection of an unresolved source and an upper limit (99.9\%) on the 
isotropic VHE \gr luminosity of 
\begin{equation}
L_\gamma (>200 \textrm{GeV})< 9.6 \times 10^{40} \textrm{ erg s}^{-1}.
\end{equation} has been derived \citep{mio_icrc}. Extended emission would not be expected given that the 
angular resolution of H.E.S.S. exceeds the angular size of the host galaxy ($\sim$7 arcmin, \cite{ned_diam}). 
The given upper limit assumes a spectral index of $\Gamma=-2.6$ for the photon spectrum.

\begin{figure}[htb]
\begin{center}
  \includegraphics[width=1\linewidth]{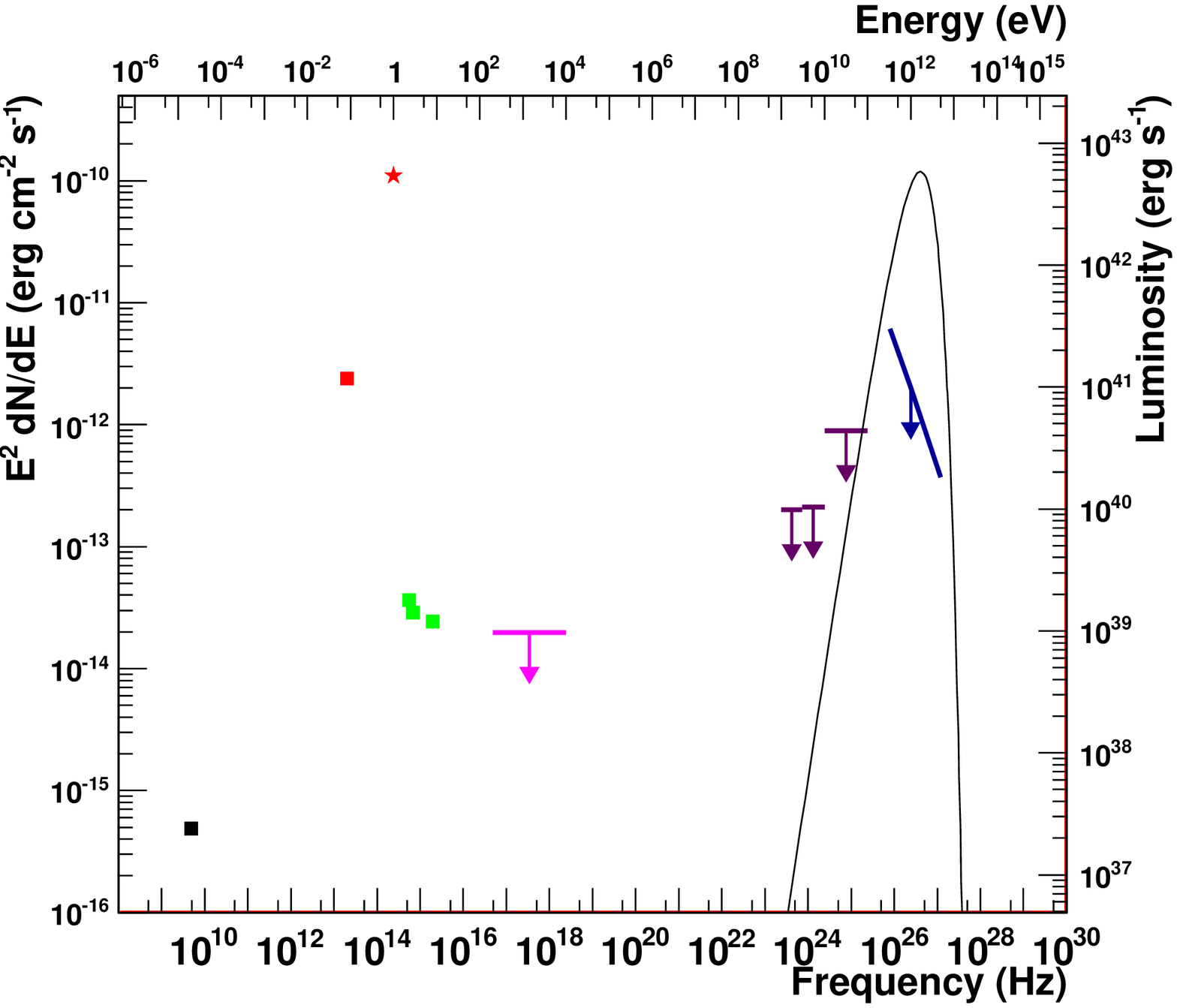}
  \caption{\small The SED of NGC 1399. All archival data are for the core region. The archival points are: 
    VLA radio data (black; \cite{sadl}; core component $<$0.4''); 15$\mu$m ISO IR data (red square; \cite{temi}; 
    5'' aperture, no host subtraction); 2MASS J-band data (red star; \cite{2mass}; 4'' aperture, no host subtraction); 
    HST optical data (green; \cite{oconn}; 0.2'' aperture), and Chandra X-ray upper limit (solid line; \cite{lowen};
    3'' aperture). The purple upper limits (95\%) are the Fermi data presented in this paper. The blue line is the 
    H.E.S.S. spectrum 2$\sigma$ upper limit \citep{mio_icrc}. The thin black line represents a toy curvature 
    spectrum, i.e., a delta-function population with energy as in eq.~(\ref{eq:cutoff}) and with maximum luminosity 
    from eq.~(\ref{eq:luminosity_adaf}) assuming a fully-developed gap.\normalsize}
 \label{fig:SED}
\end{center}
\end{figure}

While the H.E.S.S. upper limits refers to the integrated flux above 200 GeV, lower-energy gamma-ray emission 
can be studied using data obtained with the LAT instrument onboard Fermi. The data-set consists of 23 months 
of public Fermi data (04 August 2008 - 07 July 2010). The data set has been analyzed with the public software 
released by the Fermi Collaboration (ScienceTools-v9r15p2-fssc-20090808). Only events between 200 MeV 
and 100 GeV were considered and standard selection cuts for point-sources (as 
outlined in the Cicerone Fermi manual) were applied. The spectral analysis has been performed through an 
unbinned likelihood fit (gtlike tool in the public software). First a likelihood fit on the entire data range was made 
with a power law functional form $dN/dE=N_0(E/E_\mathrm{p})^{-\Gamma}$, where $N_0$ is the normalization 
at the pivot energy $E_\mathrm{p}$ and $\Gamma$ is the spectral index. The object is not significant on the whole 
energy range (2.4$\sigma$). The background has been modelled as being composed from the sources in the 
first year Fermi Catalog and from two diffuse backgrounds. The diffuse backgrounds (both galactic and extragalactic) 
have been modelled from the ones supplied along with the public software. Upper limits have been derived at a 
95\% level for the three energy bins that are of most interest for the models investigated in this paper. The assumed 
spectral index for the upper limits is $\Gamma=2$. The bins are 1-3GeV, 3-10GeV, 10-100GeV. Results are shown 
in Fig.~\ref{fig:SED}. 

\subsection{Potential Absorption of VHE $\gamma$-rays}
The absence of gamma-ray signals might, in principle, result from an annihilation of gamma-ray emission along 
the line-of-sight towards the observer. In the case of NGC 1399, however, any such absorption is unlikley:
The cross section $\sigma_{\gamma\gamma}$ of this process depends on the product of the energies of the 
colliding photons. For VHE photons, the most effective interaction at the peak of the cross section is with 
background photons of energy
$$\epsilon \approx 1\left(E/\mathrm{1TeV}\right)^{-1} \textrm{ eV.} $$ 
The optical depth resulting from this absorption, in a source of luminosity $L$ and radius $R$, is given by
\begin{equation}
\tau\left(E,R\right) \simeq \frac{L\left(\epsilon\right)\sigma_{\gamma\gamma}}{4\pi R \epsilon c}.
\end{equation}
For NGC 1399, the visibility of a 200 GeV photon would require $L\left(\epsilon=5\mathrm{eV}\right) <4 \times 
10^{41}(R/100~r_s)$ erg/s. This condition is satisfied (Fig.~\ref{fig:SED}) since the flux measured within a small 
aperture is very low. If, instead, more energetic hard photons are considered, $E > \mathrm{1 TeV}$, the 
constraints are placed by photons in the near-infrared regime $\left(\epsilon=1\mathrm{eV}\right)$. 
In the absence of measurements with small apertures, (weak) constraints can only be derived from the fairly high 
near-infrared flux obtained with an aperture which is likely to be dominated by starlight. This would set a lower limit 
on the size of the emitting region of $R\sim3\times10^3 r_{\mathrm{S}}$ in order to allow VHE photons to escape.

An additional photon field that might possibly interact with VHE \grs produced near the SMBH is provided by 
synchrotron radiation emitted by pairs \citep{lukasz-abs}. However, the critical luminosity for this to occur is 
well above the estimated VHE output, so that this process is unlikely to limit the escape of VHE photons.

Lastly, absorption on the diffuse extragalactic background light does not cause any significant attenuation due 
to the proximity of NGC 1399.

We conclude that photon-photon pair absorption is expected to be low throughout the gamma-ray range 
considered here and should not significantly affect any potential VHE emission emitted from the vicinity of the 
SMBH of NGC 1399.

\section{Discussion and Conclusion}
If efficient gap-type particle acceleration and curvature emission occurs in NGC 1399, VHE $\gamma$-ray 
emission should have been detected by Fermi and/or \hess According to eq.~(\ref{eq:luminosity_adaf}), the 
non-detection of NGC 1399 thus either suggests that (i) the strength of the ordered, magnetospheric field 
component is only a small fraction ($\leq 0.1$) of the disk magnetic field value estimated above, (ii) on 
average only a small fraction of the gap ($[h/r_g]^2\ll1$) is available for particle acceleration and/or (iii) the 
charge density $n_e$ in the vicinity of the black hole is much smaller than the Goldreich-Julian density, i.e. 
$(n_e/n_{\rm GJ})\ll1$. 

Option (i) would be incompatible with the jet power $L_{\rm jet} \simeq 10^{42}$ erg/s \citep{shurkin} being 
provided by a Blandford-Znajek-type process. Option (ii) would require that the charge density around 
the black hole exceeds the Goldreich-Julian density $n_{\rm GJ}\simeq 2 \times 10^{-3} (B/10^3\mathrm{G})$ 
cm$^{-3}$. This could be the case if efficient pair production occurs near to the black hole
\citep[e.g.,][]{moc11,lev11}. The accretion rate for NGC 1399 inferred above \citep[cf. also][]{narayan02} seems indeed close 
to the critical value $\dot{m}_c \simeq 2 \times 10^{-4}$ where annihilation of MeV photons in a two-temperature 
ADAF could lead to the injection of seed charges with density $n_e \geq n_{\rm GJ}$ \citep{lev11}. If the 
accretion rate would be sufficiently high, i.e., $\dot{m} >\dot{m}_c$ ensuring $n_e > n_{\rm GJ}$, a substantial
part of the gap is expected to be screened, suggesting $(h/r_g)^2 \ll 1$. This could make the anticipated VHE 
output, eq.~(\ref{eq:luminosity_adaf}), consistent with the VHE upper limits derived above. However, it would
also imply that the available potential, eq.~(\ref{V_effective}), is reduced down to a level where UHE proton 
acceleration will no longer be possible. On the other hand, if the accretion would be such that $\dot{m} <
\dot{m}_c$, implying $n_e < n_{\rm GJ}$ (option [iii]), fully-developed gaps ($h\sim r_g$) may exist. This could 
again make $L_{\gamma}$ consistent with the observationally inferred VHE upper limits. However, an additional
plasma source would then be needed to establish the force-free outflow believed to be present on larger scales. 
One plausible scenario relates to pair cascade formation in charge-starved magnetospheric regions due to the 
absorption of inverse Compton up-scattered photons \citep{lev11}. When applied to NGC 1399, the estimated 
multiplicity appears indeed sufficiently high ($M \simgt 10^3$) to allow the pair density to approach the 
Goldreich-Julian density. As the optical depth for pair production across the gap is larger than unity even if the 
gap is not fully restored (i.e., for $h<r_g$), cascade formation can occur on scales $< r_g$ and thereby limit 
the potential gap size. Detailed modelling would be required to self-consistently calculate the gap size and 
associated VHE emission. On the other hand, the estimated VHE \gr output $L_{\gamma}$ is roughly a fraction 
$(h/r_g)^2$ of the maximum Blandford-Znajek jet power $L_\mathrm{BZ}$. Assuming $L_{\rm BZ}=L_{\rm jet}$, 
the VHE flux upper limits imply that $(h/r_g)^2 \simlt 0.1$, again yielding conditions not conducive for efficient 
proton acceleration beyond $4 \times 10^{18}$ eV.

Nearby, passive supermassive black holes are often believed to be prime candidates for magnetospheric, 
gap-type particle acceleration and emission scenarios. The non-detection of VHE gamma-ray emission in 
NGC 1399 by current VHE instruments now suggests that this object is unlikely to be an efficient UHECR 
proton accelerator.

\acknowledgments
\small
This work has been supported by the International Max Planck Research School (IMPRS) for Astronomy \& 
Cosmic Physics at the University of Heidelberg.\\
Discussions with F. Aharonian, G. Bicknell and J. Kirk, and constructive comments by an anonymous referee 
that helped to improve the presentation are gratefully acknowledged.\\
This research has made use of the NASA/IPAC Extragalactic Database (NED) which is operated by the Jet 
Propulsion Laboratory, California Institute of Technology, under contract with the National Aeronautics and 
Space Administration.

\end{document}